\documentclass{pasj00}

\begin{document}
\SetRunningHead{Doi et al.}{VLBI observations of the most radio-loud NLS1}
\Received{2006/2/28}
\Accepted{2006/8/24}

\title{VLBI observations of the most radio-loud, narrow-line quasar SDSS~J094857.3+002225}



%
 \author{%
   Akihiro \textsc{Doi},\altaffilmark{1,2,3}
   Hiroshi \textsc{Nagai},\altaffilmark{4,3}
   Keiichi \textsc{Asada},\altaffilmark{3}
   \\
   Seiji \textsc{Kameno},\altaffilmark{5,3}
   Kiyoaki \textsc{Wajima},\altaffilmark{6}
   and
   Makoto \textsc{Inoue}\altaffilmark{3}\\}
\altaffiltext{1}{Department of Physics, Faculty of Science, Yamaguchi University,\\ Yoshida, Yamaguchi, Yamaguchi 753-8512}
\altaffiltext{2}{Department of Astronomy, Faculty of Science, The University of Tokyo,\\ 7-3-1 Hongo, Bunkyo-ku, Tokyo 113-0033, Japan}
\altaffiltext{3}{National Astronomical Observatory, 2-21-1 Osawa, Mitaka, Tokyo 181-8588, Japan}
\altaffiltext{4}{Department of Astronomical Science, The Graduate University for Advanced Studies,\\ 2-21-1 Osawa, Mitaka, Tokyo 1811-8588, Japan}
\altaffiltext{5}{Department of Physics, Faculty of Science, Kagoshima University,\\ 1-21-35 Korimoto, Kagoshima 890-0065}
\altaffiltext{6}{Korea Astronomy and Space Science Institute,\\ 61-1 Whaam-dong, Yuseong, Daejeon 305-348, Korea}



\maketitle

\begin{abstract}
We observed the narrow-line quasar SDSS~J094857.3+002225, which has the highest known radio loudness for a narrow-line Seyfert~1 galaxy~(NLS1), at 1.7--15.4~GHz with the Very Long Baseline Array~(VLBA).  This is the first very-long-baseline interferometry~(VLBI) investigation for a radio-loud NLS1.  We independently found very high brightness temperatures from (1)~its compactness in a VLBA image and (2)~flux variation among the VLBA observation, our other observations with the VLBA, and the Very Large Array~(VLA).  A Doppler factor larger than 2.7--5.5 was required to meet an intrinsic limit of brightness temperature in the rest frame.  This is evidence for highly relativistic nonthermal jets in an NLS1.  We suggest that the Doppler factor is one of the most crucial parameters determining the radio loudness of NLS1s.  The accretion disk of SDSS~J094857.3+002225 is probably in the very high state, rather than the high/soft state, by analogy with X-ray binaries with strong radio outbursts and superluminal jets such as GRS~1915+105.  
\end{abstract}

\section{Introduction}
Narrow-line Seyfert~1 galaxies (NLS1s; \cite{Osterbrock&Pogge1985,Pogge2000}) are usually found as radio-quiet objects \citep{Zhou_Wang2002}.  Various properties observed through optical and X-ray bands suggest that NLS1s are active galactic nuclei~(AGNs) of highly accreting~($L_\mathrm{bol}/L_\mathrm{Edd}\sim1$, where $L_\mathrm{bol}$ and $L_\mathrm{Edd}$ are bolometric and Eddington luminosities, respectively) systems~(e.g.,~\cite{Pounds_etal.1995,Boroson2002}).  The radio-quietness of NLS1s may be related to the suppression of radio jets emanated from accretion disks with high accretion rates \citep{Greene_etal.2006}, as well as the disks of X-ray binaries in the {\it high/soft} state (see, e.g.,~\cite{McClintock&Remillard2003} for a review).  

On the other hand, radio-loud NLS1s do exist, although they are very rare \citep{Zhou_Wang2002}.  There is no significant difference between radio-quiet and radio-loud NLS1s in optical and X-ray properties (SDSS~J094857.3+002225,~\cite{Zhou_etal.2003}; RX~J0134.2$-$4258,~\cite{Grupe_etal.2000}; RGB~J0044+193,~\cite{Siebert_etal.1999}; PKS~2004$-$447,~\cite{Oshlack_etal.2001}; PKS~0558$-$504,~\cite{Wang_etal.2001}).  The origin of the radio excess of the radio-loud NLS1s has been unknown.  Hardening of X-ray spectra during rapid X-ray flares of the radio-loud NLS1 PKS~0558$-$504 in a few minutes could arise from transient spectral dominance of synchrotron emission from relativistically boosted jets \citep{Wang_etal.2001} similar to radio-loud quasars (e.g.,~\cite{Reeves_etal.1997}).  However, properties of radio jets in NLS1s are slightly known.  Their linear sizes are $\lesssim 300$~pc \citep{Ulvestad_etal.1995}, which correspond to $\sim1$~arcsec resolutions of connected interferometries; there are only a few images that have been obtained by very-long-baseline interferometry~(VLBI) observations~\citep{Lal_etal.2004,Middelberg_etal.2004}.

\begin{table*}
\caption{Parameters of observations.\label{table1}}
\begin{center}
\begin{tabular}{lcrccc} 
\hline\hline
\multicolumn{1}{c}{Date} & Array & \multicolumn{1}{c}{$\nu_\mathrm{c}$} & $\Delta\nu$ & pol. & $t_\mathrm{scan} \times N_\mathrm{scan}$ \\
\multicolumn{1}{c}{} &  & \multicolumn{1}{c}{(GHz)} & (MHz) &  & (s) \\
\multicolumn{1}{c}{(1)} & (2) & \multicolumn{1}{c}{(3)} & (4) & (5) & (6) \\\hline
2003Oct30 & VLBA & 15.365 & 32 & L & 46$\times$52 \\
 &  & 8.421 & 32 & R & 35$\times$18 \\
 &  & 4.987 & 32 & L & 56$\times$7 \\
 &  & 2.271 & 32 & R & 85$\times$6 \\
 &  & 1.667 & 32 & L & 94$\times$6 \\
2005May09 & VLBA & 1.667 & 32 & L & 46$\times$12 \\
2003Dec15 & VLA-B & 14.94 & 100 & dual & 43$\times$2 \\
 &  & 8.460 & 100 & dual & 33$\times$1 \\
 &  & 4.860 & 100 & dual & 33$\times$1 \\
 &  & 1.425 & 100 & dual & 43$\times$1 \\\hline
\end{tabular}
\end{center}
{\footnotesize 
Col.~(1) observation date; Col.~(2) array; Col.~(3) center frequency; Col.~(4) bandwidth; Col.~(5) polarization.  L and R denote left and right circular polarization, respectively; Col.~(6) scan lengths $\times$ number of scans.
}
\end{table*}

We started VLBI-imaging several NLS1s.  SDSS~J094857.3+002225 was discovered at $z=0.584$ in the Sloan Digital Sky Survey (SDSS) early data release \citep{Williams_etal.2002}.  With FWHM(H$\beta$)$\approx$1500~km~s$^{-1}$ and undetected [OIII] lines, this source was consistent with the conventional definition of NLS1s \citep{Zhou_etal.2003}.  Its radio counterpart was found in the Faint Images of the Radio Sky at Twenty-centimeter~(FIRST) survey \citep{Becker_etal.1995}.  This has been the most radio-loud~($R\approx 2000$) object in NLS1s \citep{Zhou_etal.2003}, where $R$ is ``radio loudness'' conventionally defined as the ratio of 6~cm radio to {\it B}-band flux densities with a threshold of $R=10$ separating radio-loud and radio-quiet objects (e.g.,~\cite{Visnovsky_etal.1992,Stocke_etal.1992,Kellermann_etal.1994}).  \citet{Zhou_etal.2003} suggested that its rapid ($\sim$years) radio-flux variations were caused by relativistic boosting of a jet.  This is appropriate for the first target of VLBI observations, because of the brightest ($\sim300$~mJy) NLS1 in radio bands.  In the present paper, we report VLBI observations in milliarcsecond~(mas) resolutions for SDSS~J094857.3+002225; this is the first VLBI study for a radio-loud NLS1.  Throughout this paper, a flat cosmology is assumed, with $H_0=71$~km~s$^{-1}$~Mpc$^{-1}$, $\Omega_\mathrm{M}=0.27$, and $\Omega_\mathrm{\Lambda}=0.73$ \citep{Spergel_etal.2003}.  Therefore, 1~mas corresponds to 6.6~pc at the distance to SDSS~J094857.3+002225.  Its optical magnitude $m_B \mathrm{(SDSS)} \approx 18.89$~mag \citep{Zhou_etal.2003} corresponding to $M_B=-23.8$~mag implies that this source is a quasar rather than a Seyfert galaxy according to the standard definition of a quasar, $M_B<-23$~mag.

\section{Observations and data reduction}\label{section:observation}
We observed SDSS~J094857.3+002225 using ten stations of the Very Long Baseline Array~(VLBA) on October~30, 2003 and May~09, 2005.  The observation parameters are shown in Table~\ref{table1}.  Standard procedures were applied for data reduction using the Astronomical Image Processing System (AIPS; \cite{Greisen2003}).  Amplitude calibration used a priori gain values, together with system temperatures measured during the observations, and was typically 5\% accurate.  Imaging, deconvolution, and self-calibration for phase and amplitude were performed using the Difmap software~\citep{Shepherd1997}.  The final self-calibration in phase was done at all the frequencies with a solution interval of 10~s, which is much less than coherence times.      

SDSS~J094857.3+002225 was also observed in a snapshot mode with the Very Large Array (VLA) B-array configuration on December~15, 2003.  The VLA observations were carried out in good weather conditions.  Data were calibrated using the AIPS, according to the standard manner.  Maps in Stokes~I were made, and deconvolution and self-calibration were performed in the AIPS.  A solution interval of 10~s, similar to the VLBA data reduction, was used in self-calibration to correct of visibility phases.  The flux-scaling factors of visibility amplitude were determined by observations from the standard calibrator 3C~48, with an uncertainty of $\sim$3\%.

\section{RESULTS}\label{section:result}
\subsection{VLBA images}\label{section:resultVLBA}
In the VLBA data analyses, source structure models were established by visibility-based model fitting with point and/or elliptical-Gaussian profile(s) in the Difmap.  Here, we describe the results of the fitting, particularly at 15~GHz, where the highest spatial resolution was available during the observations.  The free parameters were flux density and two-dimensional position for a point source model, along with major axis, axis ratio, and position angle of the axis for an elliptical-Gaussian model.

\begin{figure*}
\includegraphics[width=\linewidth]{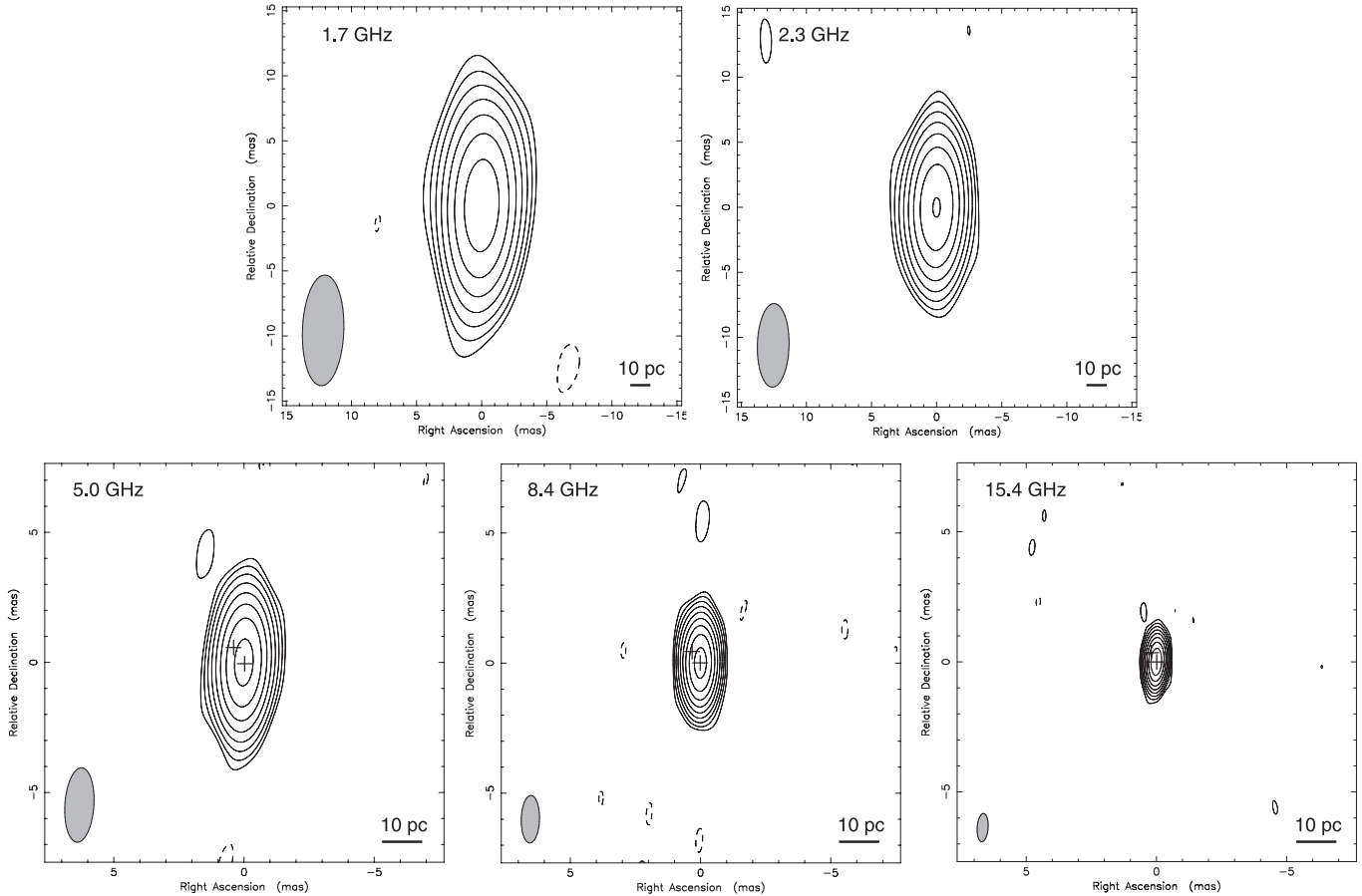}
\caption{VLBA contour maps of SDSS~J094857.3+002225 on Oct~30, 2003.  The contour levels are $3\sigma \times -1$, 1, 2, 4, 8, 16, 32, 64, 128, and 256, where $\sigma$ is the RMS of image noise at each frequency.  In the images at 5.0, 8.4, and 15.4~GHz, ``$+$'' symbols represent the positions of C0 and C1 components, which were identified by visibility-based model fitting (Section~\ref{section:resultVLBA}).}
\label{figure1}
\end{figure*}

One point source resulted in $\chi^2/dof=77821/57109$; adding one more point source showed $\chi^2/dof=73870/57106$: a dramatic improvement of $\Delta \chi^2/\Delta \rm{dof}=3950/3$, which is significant at $>$99.9\% confidence.  We found both a strong component ``C0'' and a weak component ``C1,'' not only at 15~GHz but also at 5 and 8.4~GHz.  However, the two components were not separable from each other at 1.7 and 2.3~GHz, where the resolution was poorer (Table~\ref{table2} and Fig.~\ref{figure1}).  No other significant emission in the residual images was found.  We had initially suspected that the weaker component was created by a deconvolution error, because the separation was the same as or smaller than the synthesized beams.  However, a number of test images from fake visibility data created by Monte-Carlo simulations with an identical ($u,v$)-coverage and a similar sensitivity indicate that a strong component and a weak one that are separated by $>$0.10~mas at $PA=$\timeform{30D} can be clearly resolved from each other.  We are confident, therefore, in the existence of the two components separated by 0.41~mas at 15~GHz.  Although the separations between C0 and C1 were different at observing frequencies, the position angles of C1 from C0 were consistent (Table~\ref{table2}).  

Furthermore, we attempted to fit the visibility data with two elliptical Gaussians.  However, this attempt failed because the source sizes were too small.  To evaluate how a large source can be resolved in the VLBA observation, we made imaging simulations in the same manner using fake visibility data that were made by combining a strong circular-Gaussian source of 461~mJy with various diameters and a weak point source of 21~mJy separated by 0.4~mas at $PA \sim$\timeform{30D} (see similar analyses in \cite{Taylor_etal.2004}).  
These tests suggest that the stronger source can be clearly resolved if the given size is $>$0.05~mas in diameter.  The VLBA observation should have the ability to resolve a size of $\sim$0.10~mas for the stronger component.  The size of 0.10~mas corresponds to a decrease of $\sim10$\% in visibility amplitude at the longest baseline of the VLBA at 15~GHz.  Because of the sufficiently good sensitivity and calibration, we were able to determine the fringe visibility to 5\% (see similar discussion by \cite{Kellermann_etal.1998}).   Thus, we assume the effective resolution of our VLBA observation at 15~GHz is around 0.10~mas, which is about quarter of the synthesized beam.  The component C0 is presumably compact with a size of $<0.10$~mas, corresponding to $<$0.66~pc on a linear scale.

\subsection{VLA images}\label{section:resultVLA}
In the VLA images, only a single, unresolved component was found at all the frequencies, except for 1.4~GHz, where radio interference affected the data.  No diffuse emission component was found, even in ($u,v$)-tapered images.  Flux density was measured by Gaussian model fitting on the image domain using the task JMFIT of the AIPS.

\begin{table*}
\caption{Observational results.\label{table2}}
\begin{center}
\begin{tabular}{lcrclcllccc} 
\hline\hline
\multicolumn{1}{c}{$\nu$} & Comp. & \multicolumn{3}{c}{$S_\nu$} & $\sigma_\mathrm{rms}$ & \multicolumn{1}{c}{$\theta_\mathrm{maj}$} & \multicolumn{1}{c}{$\theta_\mathrm{min}$} & $PA$ & $l$(C0--C1) & $PA$(C0--C1) \\
\multicolumn{1}{c}{(GHz)} &  & \multicolumn{3}{c}{(mJy)} & (mJy beam$^{-1}$) & \multicolumn{1}{c}{(arcsec)} & \multicolumn{1}{c}{(arcsec)} & (deg) & (mas) & (deg) \\
\multicolumn{1}{c}{(1)} & (2) & \multicolumn{3}{c}{(3)} & (4) & \multicolumn{1}{c}{(5)} & \multicolumn{1}{c}{(6)} & (7) & (8) & (9) \\\hline
\multicolumn{5}{l}{{\bf VLBA observed on Oct 30, 2003}} &  & \multicolumn{1}{c}{} & \multicolumn{1}{c}{} &  &  &  \\
15.365L & C0 & 461.3  & $\pm$ & 23.1  & 0.65  & 0.0011  & 0.0004  & $-3.3$ & \ldots & \ldots \\
 & C1 & 20.6  & $\pm$ & 1.6  & \ldots & \ldots & \ldots & \ldots & 0.41$\pm$0.02 & $+$30.7 \\
8.421R & C0 & 363.0  & $\pm$ & 18.2  & 0.37  & 0.0019  & 0.0008  & $-2.7$ & \ldots & \ldots \\
 & C1 & 25.2  & $\pm$ & 1.4  & \ldots & \ldots & \ldots & \ldots & 0.51$\pm$0.01 & $+$34.9 \\
4.987L & C0 & 258.9  & $\pm$ & 13.0  & 0.56  & 0.0031  & 0.0012  & $-3.8$ & \ldots & \ldots \\
 & C1 & 32.0  & $\pm$ & 2.4  & \ldots & \ldots & \ldots & \ldots & 0.75$\pm$0.02 & $+$33.1 \\
2.271R & \ldots & 245.4  & $\pm$ & 12.3  & 0.67  & 0.0072  & 0.0029  & $-2.8$ & \ldots & \ldots \\
1.667L & \ldots & 173.0  & $\pm$ & 8.7  & 0.55  & 0.0098  & 0.0040  & $+0.9$ & \ldots & \ldots \\
\multicolumn{5}{l}{{\bf VLBA observed on May 09, 2005}} &  &  &  &  &  &  \\
1.667L & \ldots & 85.7  & $\pm$ & 4.4  & 0.58  & 0.0104  & 0.0039  & $-1.4$ & \ldots & \ldots \\
\multicolumn{5}{l}{{\bf VLA-B observed on Dec 15, 2003}} &  &  &  &  &  &  \\
14.94 & \ldots & 334.1  & $\pm$ & 10.1  & 0.56  & 0.9  & 0.4  & $-48.1$ & \ldots & \ldots \\
8.460 & \ldots & 359.8  & $\pm$ & 10.8  & 0.36  & 1.8  & 0.7  & $-50.3$ & \ldots & \ldots \\
4.860 & \ldots & 304.7  & $\pm$ & 9.7  & 1.89  & 3.0  & 1.2  & $-49.9$ & \ldots & \ldots \\
1.425 & \ldots & 176.6  & $\pm$ & 6.0  & 1.53  & 9.3  & 4.2  & $-50.6$ & \ldots & \ldots \\\hline
\end{tabular}
\end{center}
{\footnotesize 
Col.~(1) observing frequency.  L and R denote left and right circular polarization, respectively; Col.~(2) component name.  C0 and C1 denote the bright and weaker component, respectively, which were spatially resolved by visibility-based model fitting~(Section~\ref{section:resultVLBA}); Col.~(3) flux density; Col.~(4) RMS of image noise; Col.~(5) FWHM of synthesized beam at major axis; Col.~(6) at minor axis; Col.~(7) position angle of synthesized beam; Col.~(8) separation between C0--C1; Col.~(9) position angle of C1 from C0.
}
\end{table*}

\subsection{Measured flux densities}\label{section:resultflux}
Measured flux densities are listed in Table~\ref{table2}.  We estimated the total error from the root sum square of the uncertainty of amplitude calibration and the error in Gaussian fitting (including thermal noise).  

The total spectra of observed VLBA flux densities were slightly inverted ($\alpha>0$, $S_\nu \propto \nu^{+\alpha}$), which is consistent with past observations.  Non-simultaneous observations from 111.5~mJy at 1.4~GHz by the FIRST survey~\citep{Becker_etal.1995} and 234~mJy at 8.4~GHz by the Jodrell Bank VLA Astrometric Survey~\citep{Browne_etal.1998} showed an inverted spectrum.  Simultaneous observations at 2.7, 5, and 10.7~GHz~\citep{Reich_etal.2000} showed a spectral index, $\alpha=-0.24\pm0.08$.  Our VLA flux densities were consistent with or less than those of the VLBA; this suggests that there is no missing flux in the VLBA images of this object.

The VLBA flux density at 1.7~GHz decreased to 50\% in 19~months.  The flux density at 15~GHz also decreased to 69\% in 46~days between the VLBA and VLA observations.  We believe that these were intrinsic decreases of flux density rather than losses of high-frequency coherence on the VLA data.  The VLA observations were made in good weather conditions.  The coherence loss should be inhibited to the same level as that of the VLBA, since a sufficiently short solution interval, 10~s, was adopted for phase self-calibration to both the VLBA and VLA data (Section~\ref{section:observation}).

\section{Discussion}\label{section:discussion}

\subsection{Very high brightness temperatures}\label{section:brightnesstemperature}

We can derive brightness temperatures from both the VLBA image and the flux variability.  Using imaging analysis, brightness temperature in the rest frame is calculated using (cf. \cite{Lahteenmaki_etal.1999})
\begin{equation}
T_\mathrm{Brest}(\mathrm{img})=\frac{1+z}{\delta} \frac{c^2 S_{\nu\mathrm{obs}}^\mathrm{obs}}{2k_\mathrm{B}\nu_\mathrm{obs}^2 (\theta_\mathrm{obs} / 2)^2 \pi},
\label{equation:TB}
\end{equation}
where $c$ is the speed of light, $k_\mathrm{B}$ is the Boltzmann constant, $S_\nu$ is the flux density at frequency $\nu$, $\theta$ is the angular diameter of a source, and suffixes of ``obs'' and ``rest'' denote quantities in the observer frame and the rest frame, respectively.  The Doppler factor is defined as $\delta \equiv \sqrt{1-\beta^2}/(1-\beta\cos{\phi})$, where $\beta \equiv v/c$ ($v$ is the source speed), and $\phi$ is the ``viewing angle'' between the direction of the source velocity and our line of sight.  Given an apparent diameter of $<$0.10~mas for C0 in the 15-GHz VLBA data (Section~\ref{section:resultVLBA}), $T_\mathrm{Brest}(\mathrm{img})>5.5\times10^{11}/\delta$~K.  

Alternatively, with flux variation, a brightness temperature in the rest frame can be calculated independently using (cf. \cite{Lahteenmaki_etal.1999})
\begin{equation}
T_\mathrm{Brest}(\mathrm{var}) > \frac{1+z}{\delta^3} \frac{D_\mathrm{co}^2 \Delta S_{\nu_\mathrm{obs}}^\mathrm{obs}}{2 \pi k_\mathrm{B} \nu_\mathrm{obs}^2 (\Delta t_\mathrm{obs})^2},
\label{equation:TBrest_variable}
\end{equation}
where $D_\mathrm{co}$ is the comoving distance, $\Delta S_{\nu_\mathrm{obs}}^\mathrm{obs}$ is a change of observed flux density in a period, of $\Delta t_\mathrm{obs}$.  In this equation, we assume exponential decay as seen in radio outbursts in blazars \citep{Valtaoja_etal.1999}.  We found that the flux density decreases by 31\% in 46~days at 15~GHz, and by 50\% in 19 months at 1.7~GHz (Section~\ref{section:resultflux}), which implies that $T_\mathrm{Brest}(\mathrm{var})>3.3\times10^{13}/\delta^3$~K and $T_\mathrm{Brest}(\mathrm{var})>1.1\times10^{13}/\delta^3$~K, respectively.  A very high brightness temperature in this source has also been suggested by \citet{Zhou_etal.2003} from 1.4-GHz flux variation in $\sim$years time scale.    These very high brightness temperatures form clear evidence for a nonthermal process in the central engine of this NLS1.

\subsection{Highly relativistic jets}\label{section:relativisticjet}

Based on near equipartition of energies between radiating particles and a magnetic field, the brightness temperature of nonthermal radio emission should be within a physically realistic upper limit of $\sim2\times10^{11}$~K in the rest frame \citep{Readhead1994}.  Therefore, the Doppler factor should be $\delta>2.7$--5.5 for the observed brightness temperatures (Section~\ref{section:brightnesstemperature}).  The observed radio emissions are presumably from highly relativistic jets and are enhanced by Doppler boosting.  The jets should approach us with a viewing angle of $\theta<$\timeform{22D} and a speed of $\beta>0.76$ for $\delta>2.7$ (Section~\ref{section:brightnesstemperature}).  C1, spatially resolved in the VLBA data (Section~\ref{section:resultVLBA}), may be one of the expanding jet components.  

\begin{figure}
\includegraphics[width=1.0\linewidth]{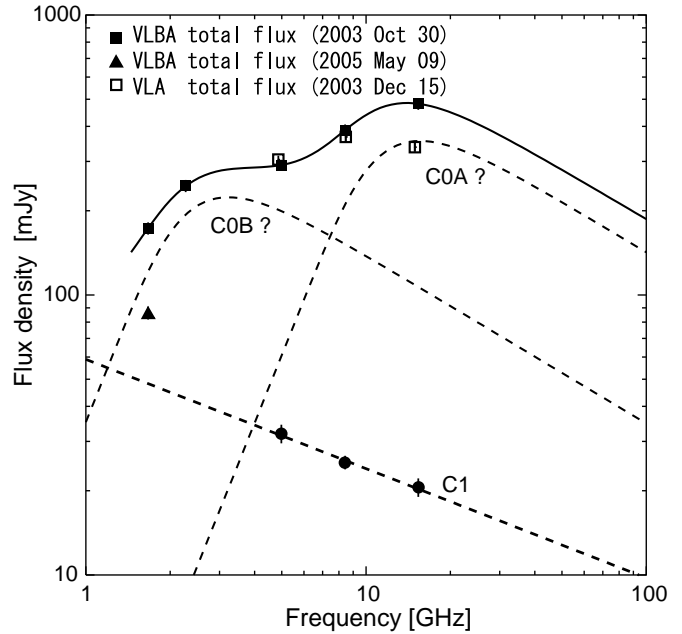}
\caption{Radio continuum spectra of SDSS~J094857.3+002225.  Symbols representing total flux densities are identified in top corner.  Filled circles particularly describe the flux densities for component C1.  The solid curve represents the spectral model that was determined by spectral model-fitting to the VLBA total fluxes using two self-absorbed synchrotron spectra (``C0A'' and ``C0B'' as dashed curves) with an intrinsic spectral index of $\alpha_0=-0.6$ and one simple power-law spectrum (C1 as a dashed line).  The spatially single component C0 was divided into two spectrally separate components, C0A and C0B with peak frequencies of $\sim$16 and 3~GHz, respectively.  See Section~\ref{section:relativisticjet} for details.}
\label{figure2}
\end{figure}

Because of the very high brightness temperature, synchrotron self-absorption could be a possible explanation for the inverted spectrum as observed.  We show an example of spectral model fitting to the observed VLBA total fluxes using power-law spectra with a low-frequency cut-off due to synchrotron self-absorption: $S_\nu = S_{\nu_0} \nu^{2.5}[1-\exp{(-\tau\nu^{\alpha_0-2.5})}]$, where $S_{\nu_0}$ is a scaling constant, $\tau$ is the optical depth, and $\alpha_0$ is the intrinsic (not absorbed) spectral index.  Spectral model fitting was performed on the assumptions that the observed spectrum of C0 is caused by a superposition of self-absorbed synchrotron spectra of two components, and that the observed spectrum of C1 can be represented by a single power-law spectrum.  As a result, C0 consists of two convex spectra with peak frequencies at $\sim$16~GHz~(``C0A'') and $\sim$3~GHz~(``C0B''), as shown in Fig.~\ref{figure2}.  The Doppler boosting can be responsible for such high frequencies of the spectral peaks, since $\nu_\mathrm{obs} = [\delta/(1+z)] \nu_\mathrm{rest}$.  The intrinsic peak frequencies could be at several hundreds of MHz.  Such high-frequency peaked spectra are reminiscent of those of the Seyfert galaxy III~Zw~2 with two compact radio components and an inverted high-frequency spectrum in total flux \citep{Falcke_etal.1999}.  These components in III~Zw~2 are also highly relativistic jets \citep{Brunthaler_etal.2000}.

\subsection{Implications of SDSS~J094857.3+002225}
Our observations showed a slightly inverted spectrum at $\sim$5~GHz and $R\approx2530$, which is derived from $m_B \mathrm{(SDSS)} \approx 18.89$~mag \citep{Zhou_etal.2003} and the VLBA total flux density at 5~GHz.  Given $\delta>5.5$ (Section~\ref{section:brightnesstemperature}), the peak frequency of C0A, $\sim$16~GHz, should be $<$2.9~GHz before Doppler beaming.  In the case where $\alpha=-0.6$, a flux density should be amplified by a factor of $>$462 from an intrinsic spectrum at the same frequency.  Therefore, there is a possibility that SDSS~J094857.3+002225 is a radio-quiet object with a steep spectrum in the rest frame, like typical NLS1s \citep{Moran2000,Zhou_Wang2002,Ulvestad_etal.1995}.

Flat or inverted radio spectra have also been found in the other radio-loud NLS1s, e.g., RGB~0044+193 \citep{Siebert_etal.1999}, 2E~1640+5345, RXS~J16290+4007, and RX~J16446+2619 \citep{Zhou_Wang2002}.  Doppler boosting may be responsible for their spectra and radio loudness as well as those of SDSS~J094857.3+002225.  The observed optical and X-ray properties and black-hole masses of radio-loud NLS1s are not significantly different from those of radio-quiet ones (e.g., \cite{Siebert_etal.1999,Wang_etal.2001}).  From our observations, we suggest that the Doppler factor is a crucial parameter determining the radio loudness of NLS1s: at least some of NLS1s could possess the ability to generate highly relativistic jets.

Analogies have been drawn between X-ray binaries and AGNs (e.g., \cite{Maccarone_etal.2003}).  NLS1s are thought to be AGNs with high accretion rates.  Although X-ray binaries in the {\it high/soft} state with accretion rates of $L/L_\mathrm{Edd}\sim0.03$--0.1 have no relevance to jets, strong radio flares have been seen during the {\it very high} state, which is thought to be a disk in $L/L_\mathrm{Edd}>0.1$, such as GRS~1915+105, with highly relativistic jets (e.g., \cite{Fender2003}).  SDSS~J094857.3+002225 has the potential to be compatible with the {\it very high} state, rather than the {\it high/soft} state.  Radio-loud NLS1s will be appropriate targets to investigate for the {\it very high} state on super-massive black hole systems.

\section{Summary}
Our VLBA and VLA observations have revealed that SDSS~J094857.3+002225, which shows the highest known radio loudness for an NLS1, has (1)~significant variability on time scales of 46~days and 19~months, (2)~an inverted spectrum, and (3)~multiple components with a size $<$0.10~mas.  The derived brightness temperatures were significantly higher than the intrinsic limit, implying Doppler-beaming with a Doppler factor $\delta>2.7$--5.5 and, therefore, relativistic jets.  At least some NLS1s could be able to generate highly relativistic jets.  We suggest that the Doppler factor is a crucial parameter determining the radio loudness of NLS1s.  On the analogy of X-ray binaries, SDSS~J094857.3+002225 is probably equivalent to an AGN in the {\it very high} state rather than the {\it high/soft} state, because of the highly relativistic jets.

\bigskip


The VLBA and VLA are operated by the National Radio Astronomy Observatory~(NRAO), a facility of the National Science Foundation operated under cooperative agreement by Associated Universities, Inc.  This research has made use of NASA Astrophysics Data System~(ADS) bibliographic services.  We thank the referee for suggestions that helped considerably in clarifying the paper.


\end{document}